\documentclass[sigconf]{acmart}

\usepackage{amsmath}
\usepackage{balance}

\AtBeginDocument{%
  }

\copyrightyear{2025}
\acmYear{2025}
\setcopyright{acmlicensed}
\acmConference[KDD '25]{Proceedings of the 31st ACM SIGKDD Conference on Knowledge Discovery and Data Mining V.2}{August 3--7, 2025}{Toronto, ON, Canada}
\acmBooktitle{Proceedings of the 31st ACM SIGKDD Conference on Knowledge Discovery and Data Mining V.2 (KDD '25), August 3--7, 2025, Toronto, ON, Canada}
\acmDOI{10.1145/3711896.3737264}
\acmISBN{979-8-4007-1454-2/2025/08}

\begin{document}
% \newtheorem{theorem}{Theorem}[section]
% \newtheorem{corollary}[theorem]{Corollary}
% \newtheorem{dfn}{Definition}
% \newtheorem{hypothesis}{Hypothesis}
% \newtheorem{lemma}[theorem]{Lemma}
% \newtheorem{proposition}[theorem]{Proposition}
% \newtheorem{remark}{Remark}
% \newtheorem{example}{Example}

\title{Supervised Learning-enhanced Multi-Group Actor Critic for Live Stream Allocation in Feed}

\author{Jingxin Liu}
\authornote{Corresponding author.}
%%\authornote{Both authors contributed equally to this research.}
%\email{webmaster@marysville-ohio.com}
\affiliation{%
  \institution{Kuaishou Technology}
  \city{Beijing}
  \country{China}
}
\email{liujingxin05@kuaishou.com}

\author{Xiang Gao}
\affiliation{%
  \institution{Kuaishou Technology}
  \city{Beijing}
  \country{China}}
\email{gaoxiang12@kuaishou.com}

\author{YiSha Li}
\affiliation{%
  \institution{Kuaishou Technology}
  \city{Beijing}
  \country{China}}
\email{liyisha@kuaishou.com}

\author{Xin Li}
\affiliation{%
  \institution{Kuaishou Technology}
  \city{Beijing}
  \country{China}
}
\email{lixin05@kuaishou.com}

\author{Haiyang Lu}
\affiliation{%
 \institution{Kuaishou Technology}
 \city{Beijing}
 \country{China}
}
\email{luhaiyang@kuaishou.com}

\author{Ben Wang}
\affiliation{%
  \institution{Kuaishou Technology}
  \city{Beijing}
  \country{China}
}
\email{benwang177@gmail.com}


\renewcommand{\shortauthors}{Jingxin Liu et al.}
%% No italics, no superscripts
%% Use footnote or author note to identify equal contribution and/or contact author info

\begin{abstract}
In the context of a short video \& live stream mixed recommendation scenario, the live stream recommendation system (RS) decides whether to allocate at most one live stream to the video feed for each user request. The inappropriate policy which ignores the long-term negative impact of live stream allocation can significantly affect app usage duration and user retention. To maximize long-term user engagement, it is crucial to determine an optimal policy for accurate live stream allocation. Recently, reinforcement learning (RL) has been widely applied in recommendation systems to capture long-term user engagement. However, traditional RL algorithms often face divergence and instability problems, which restricts application and deployment in large-scale industrial recommendation systems, especially in the aforementioned challenging scenario. To address these challenges, we propose a novel \textbf{S}upervised \textbf{L}earning-enhanced \textbf{M}ulti-\textbf{G}roup \textbf{A}ctor \textbf{C}ritic algorithm (SL-MGAC). Specifically, we introduce a supervised learning-enhanced actor-critic framework that incorporates \textit{variance reduction} techniques, where multi-task supervised reward learning helps restrict bootstrapping error accumulation during critic learning. Additionally, we design a multi-group state decomposition module for both actor and critic networks to reduce prediction variance and improve model stability. We also propose a novel reward function to prevent overly greedy live stream allocation. Empirically, we evaluate the SL-MGAC algorithm using offline policy evaluation (OPE) and online A/B testing. Experimental results demonstrate that the proposed method not only outperforms baseline methods under platform-level constraints, but also exhibits improved stability in online recommendation scenarios.
\end{abstract}





\maketitle

\begin{figure}[htbp]
\centering
\includegraphics[width=0.45\textwidth]{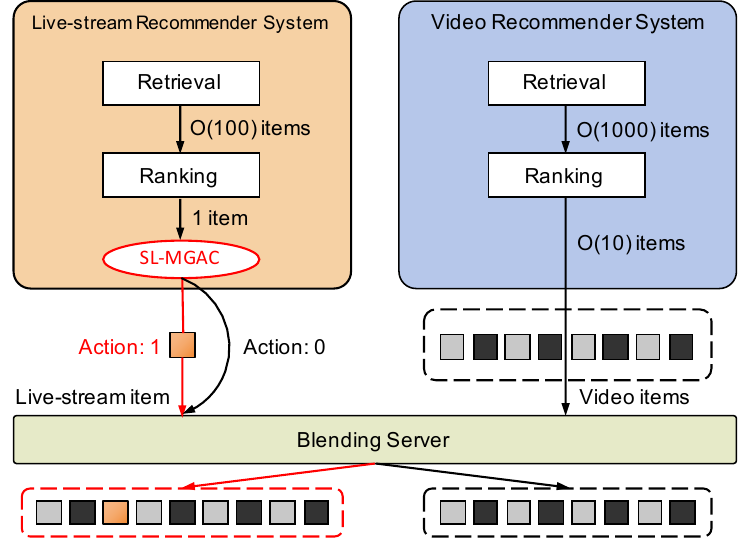}
\caption{Structure of a short video \& live stream mixed recommendation system(RS). The decision making of SL-MGAC takes place in the final stage of live stream RS. }
\label{fig:1}
\end{figure}

\section{Introduction}
% Background
We consider a challenging sequential decision making scenario in a short video \& live stream blended recommendation system, as shown in Fig. \ref{fig:1}. The system consists of three components: a live streaming recommendation system, a short video recommendation system, and a blending server. For each user request with timestamp $t$, the live stream recommendation system decides whether to inject the recommended live stream into the video feed, while the short video recommendation system suggests $B$ (with $B<10$) videos. The blending server then mixes and rearranges the single live stream with the $B$ videos to form the final recommendation list. Our objective is to find a personalized optimal live stream allocation policy that not only maximizes user long-term engagement with live streams, but also meets the constraint of not causing a negative impact on app usage duration and user retention, which can be modeled as an infinite request-level Constrained Markov Decision Process (CMDP) \cite{altman2021constrained}.

Currently, deep reinforcement learning has shown great potential in learning policies to maximize long-term rewards in various research domains, such as computer vision, robotics, natural language processing, gaming, and recommendation systems \cite{afsar2022reinforcement, li2017deep}. In the context of recommendation systems, reinforcement learning is applied to optimize long-term user engagement \cite{zou2019reinforcement} and improve user retention \cite{cai2023reinforcing}. Numerous RL applications have been proposed for real-world recommendation systems, including slate recommendation \cite{ie2019slateq, deffayet2023generative, liu2023exploration}, personalized search ranking \cite{miao2021sequential}, and advertisement allocation \cite{liao2022cross}.

However, the intrinsic issues of divergence and instability associated with traditional reinforcement learning algorithms (RL) \cite{franccois2015discount, fujimoto2018addressing, kumar2019stabilizing, dasagi2019ctrl} are significantly exacerbated in the short video \& live stream mixed recommendation system. A primary possible reason for this is the drastic fluctuations in both the live stream supply scale and user interaction behaviors over time, as shown in Fig. \ref{fig:2} in Appendix \ref{Appdenx: C}. RL models often struggle to learn effective policies from data that exhibit such high variance across both time scales and user scales. In practice, we observe that RL models frequently encounter issues of \textit{policy deterioration} or \textit{model collapse} \cite{dohare2023overcoming}. Most importantly, the live-stream allocation agent is \textbf{the final module of the live-stream RS}, as shown in Fig \ref{fig:1}. If the RL agent becomes highly unstable or collapses in the online environment, it may lead to an excessive injection of live-stream content into the short-video feed. In turn, this can negatively impact the exposure of advertisements and E-commerce videos, potentially causing severe system malfunctions and significant platform losses. Hence, in the industrial RS, this RL application is \textbf{more risky} than existing RL methods \cite{cai2023reinforcing,zhang2022multi,zhang2024unex}, which are only applied in the multi-rank score aggregation of the ranking stage.

Furthermore, like advertisement, live stream is not welcomed by every user. Allocating too many live streams for a user will significantly interrupt the user's short video interest and lead to a decrease in the user's app usage duration, which will eventually affect the user engagement and user retention. Therefore, we cannot merely focus on the user feedback for a single request, but rather should optimize the long-term user experience.

% A primary possible reason for this is the drastic fluctuations in both the live stream supply scale and user interaction behaviors over time. As shown in Fig. \ref{fig:2} a, the live stream room count and live stream viewer count curves vary significantly on a daily basis, making it difficult for the RL algorithm to learn optimal policies. Furthermore, as illustrated in Fig. \ref{fig:2}b, the distributions of user interaction labels (e.g., live streaming and short video watch time) are highly noisy and exhibit large variance. RL models often struggle to learn effective policies from data that exhibit such high variance across both time and user scales. In practice, we observe that RL models frequently encounter issues of \textit{policy deterioration} or \textit{model collapse} \cite{dohare2023overcoming}.

% Recently, a growing literature has focused on RL divergence and instability \cite{mao2018variance, dohare2023overcoming, yue2024understanding}. For policy-based RL model, the main reason of instability is the high variance of gradient estimates. For value-base RL model, the main reason of instability comes from inadvertent temporal difference (TD) error accumulation in Q-value estimation. Hence, existing variance reduction works \cite{greensmith2004variance, kumar2019stabilizing, anschel2017averaged,mao2018variance, romoff2018reward} try to enhance the stability of policy gradient \cite{mao2018variance} and TD learning \cite{xu2020reanalysis}.

To address the aforementioned problems, we propose a novel \textbf{S}upervised \textbf{L}earning-enhanced \textbf{M}ulti-\textbf{G}roup \textbf{A}ctor \textbf{C}ritic algorithm, SL-MGAC. Given the significant variation in users' interest in live stream content, which introduces high variance at both the feature and feedback levels, we incorporate separate \textbf{M}ulti-\textbf{G}roup \textbf{S}tate \textbf{D}ecomposition (MG-SD) modules within both the actor and critic networks. Additionally, we combine multi-task supervised reward learning with traditional critic learning to not only restrict temporal difference (TD) error accumulation during model training but also improve the accuracy of Q-value estimation. Through experiments, we compare the SL-MGAC method with competitive baselines using offline policy evaluation (OPE) \cite{uehara2022review} and online A/B tests to demonstrate its effectiveness and stability.

The main contributions of this paper are as follows.

\begin{itemize}
\item We introduce SL-MGAC, a novel RL model that outperforms existing methods in maximizing the user engagement of live stream while satisfying the platform-level constraints on app usage duration and user retention.
\item We propose new \textit{variance reduction} \cite{greensmith2004variance, kumar2019stabilizing, anschel2017averaged,mao2018variance, romoff2018reward} techniques, including multi-group state decomposition, distribution discretization for reward learning, reward normalization, and Q value normalization.
\item We integrate multi-task supervised reward learning with traditional critic learning to alleviate TD error accumulation and improve the accuracy of Q-value estimation.
\item We successfully deploy the SL-MGAC model in a challenging short video and live stream mixed recommendation scenario for \textit{Kwai}, a short video app with over 100 million users.
\end{itemize}

\begin{figure*}[htbp]
\centering
\includegraphics[width=0.9\textwidth]{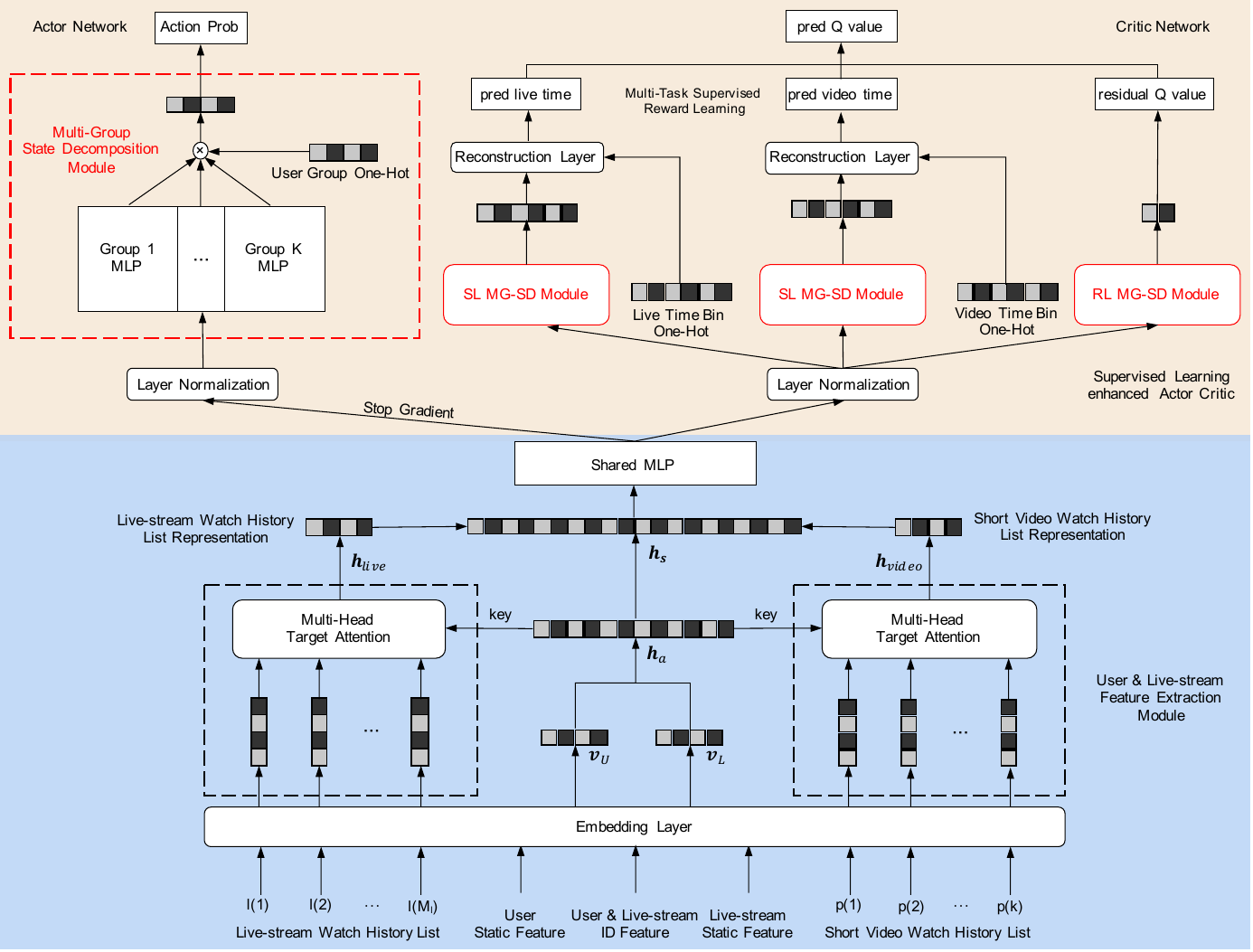}
\caption{Overall framework of the SL-MGAC algorithm. The SL (RL) MG-SD Module is short for the Multi-Group State Decomposition Module for supervised reward learning and critic learning.}
\label{fig:3}
\end{figure*}

\section{Problem Formulation}
As shown in Fig. \ref{fig:1}, in the short video \& live stream mixed recommendation system, the live stream recommendation system is treated as an agent that interacts with various users and receives user feedback over time. However, an agent considering instant user feedback of a single request may allocate live streams greedily in a short video feed and eventually affect user retention. Therefore, the optimal live stream allocation control problem is modeled as an infinite request level CMDP to maximize the cumulative live stream reward and satisfy the platform-level constraint, for example, live stream allocation cannot reduce the total app usage duration.

Formally, we define live stream allocation CMDP as a tuple of six elements $\left(\mathcal{S}, \mathcal{A}, \mathcal{P}, \mathcal{R}, \gamma, \mathcal{C} \right)$:
\begin{itemize}
\item \textbf{State space} $\mathcal{S}$: This is the set of user interaction states $s$, which includes user static features (e.g., user ID, location, gender, country), user history features (e.g., live stream watch history, short video watch history) and item features (e.g., live stream ID, author ID, author gender, etc). We limit the length of user live stream and short video history lists by timestamp order, keeping only the top 50 items.
\item \textbf{Action space} $\mathcal{A}$: The action $a \in \mathcal{A}$ represents the decision of whether to inject a recommended live stream in response to a user’s request. We define the action $a$ as a binary variable, where $a = 1$ means injecting a live stream. % and $a = 0$ means not injecting it.
\item \textbf{Transition Probability} $\mathcal{P}$: The transition probability is denoted as $p(s_{t+1}|s_t, a_t)$, determined by the environment.
\item \textbf{Reward Function} $\mathcal{R}$: The reward function $\mathcal{R}$ is a mapping from state $s_t$ and action $a_t$ at timestamp $t$, which can be formulated as $r(s_t, a_t): \mathcal{S} \times \mathcal{A} \rightarrow \mathbb{R}$. 
\item \textbf{Discount Factor} $\gamma$: The discount factor $\gamma \in [0, 1]$ is used in the calculation of the cumulative reward. Typically, we set $\gamma < 1$ during model training. 
\item \textbf{Constraint} $\mathcal{C}$: The platform-level constraint is an instantaneous constraint, that is, the live stream watch time should be longer than the average video watch time per request as much as possible. This prevents a negative impact on app usage duration, which could happen if the live stream watch time is shorter. By meeting this constraint, we ensure that live streams boost users' long-term app engagement instead of reducing their time spent on it.
\end{itemize}

Overall, the concrete reward function $r(s_t, a_t)$ and constraint function $c(s_t, a_t)$ is shown below:

\begin{equation}
\begin{aligned}
r(s_t, a_t) &= y_l \\
c(s_t, a_t) &= \frac{1}{B} y_v - y_l
\end{aligned}
\label{eq:0}
\end{equation}
where $y_l$ is the live stream watch time, $y_{v}$ is the total video watch time, and $B$ is the number of videos in a request.

The optimization objective of aforementioned optimal live stream allocation control problem is shown as follows:
\begin{equation}
\begin{aligned}
&\max_{\pi} J_R(\pi) \\
& s.t. \quad J_C(\pi) \leq \epsilon \\
\end{aligned}
\label{eq:1}
\end{equation}
where $J_R(\pi) = \mathbb{E} \left[\sum_{t = 0}^{\infty} \gamma^t r(s_t, a_t) \right]$, $J_C(\pi) = c(s_t, a_t), \forall t \in [0, \infty)$, $\pi(s_t, a_t)$ is the live stream allocation policy to be optimized. 

A common method to solve the above CMDP is to transform the problem into a min-max optimization by introducing the Lagrange multiplier $\lambda$:

\begin{equation}
\begin{aligned}
(\pi^{*}, \lambda^{*}) = \arg\min_{\lambda \geq 0} \max_{\pi} J_{R}(\pi) - \lambda (J_{C}(\pi) - \epsilon)
\end{aligned}
\label{eq:6}
\end{equation}

Inspired by the RCPO \cite{tessler2018reward} algorithm, the primal-dual optimization problem can be transformed into an equivalent unconstrained
problem with penalized reward function by substituting into $r(s_t, a_t)$ and $c(s_t, a_t)$ of Eq. \ref{eq:0}:

\begin{equation}
\begin{aligned}
\hat{r}(\lambda, s_t, a_t) &= r(s_t, a_t) - \lambda c(s_t, a_t) \\
&= y_{l} - \lambda(\frac{1}{B} y_v - y_l) \\
& = (1 + \lambda) y_l - \frac{\lambda}{B} y_v
\end{aligned}
\label{eq:7}
\end{equation}

In practice, we solve the unconstrained MDP problem with a simplified reward function as follows:
\begin{equation}
\begin{aligned}
&\max_{\pi} \mathbb{E} \left[\sum_{t = 0}^{\infty} \gamma^t\left(y_{l} - \frac{\lambda}{B} y_v\right) \right] \\
\end{aligned}
\label{eq:8}
\end{equation}
where $\lambda$ is a hyper-parameter to carefully control the final live stream allocation ratio. From the perspective of constrained reinforcement learning, the term $\frac{\lambda}{B} y_{v}$ can be viewed as an implicit cost constraint. Note that we leave the further discussion of the above reward function in the Appendix \ref{Appdenx: D}.

\section{Proposed Framework}
To address the intrinsic divergence and instability issues of reinforcement learning and successfully deploy the RL agent in our mixed short video \& live stream recommendation system, we propose a novel Supervised Learning-enhanced Multi-Group Actor-Critic algorithm (SL-MGAC), as shown in Fig. \ref{fig:3}. SL-MGAC incorporates a supervised learning-enhanced actor-critic model with a shared user \& live stream feature extraction module, along with independent Multi-Group State Decomposition (MG-SD) modules. For clarity, only one critic network is shown in Fig. \ref{fig:3}, while the remaining three critic networks, which use the same architecture, are omitted.

\subsection{User \& live stream Feature Extraction Module}
The user \& live stream feature extraction module aims to generate non-linear embedding representations for the state $s_t$. First, we use a unified embedding layer to map the user's static features, live stream features, historical live stream list features, and short video-list features into low-dimensional dense representations. We denote $\mathbf{v}_{U}$, $\mathbf{v}_{L}$, $\left\{\mathbf{e}_1, \cdots, \mathbf{e}_{M_l}\right\}$, $\left\{\mathbf{e}'_1, \cdots, \mathbf{e}'_{M_v}\right\}$ as the corresponding embedding vectors or sets of embedding vectors, respectively. We then define $\mathbf{h}_{a} = [\mathbf{v}_{U}, \mathbf{v}_{L}]$ as the concatenation of $\mathbf{v}_{U}$ and $\mathbf{v}_{L}$.

To aggregate historical live stream (short video) representations, we introduce the \textit{target attention} mechanism \cite{vaswani2017attention, zhou2018deep}, which is defined as follows:

\begin{equation}
\begin{aligned}
\mathbf{h}_{live} &= \sum_{i = 0}^{M_l} f_{l}(\mathbf{h}_{a}, \mathbf{e}_i) \mathbf{e}_{i} \\
\mathbf{h}_{video} &= \sum_{j = 0}^{M_v} f_{v}(\mathbf{h}_{a}, \mathbf{e}'_j) \mathbf{e}_{j}' \\
\end{aligned}
\label{eq:5}
\end{equation}
where $f_l$, $f_v$ are different target attention functions, such as a feed-forward network whose output is a normalized score.

After aggregating the historical live stream and short video features via two separate attention networks, we concatenate all of the above embedding vectors to form $\mathbf{h}_{s} = [\mathbf{v}_{U}, \mathbf{v}_{L}, \mathbf{h}_{live}, \mathbf{h}_{video}]$. We then use a shared multi-layer perceptron (MLP) for both the actor and critic networks to obtain the latent representation of state $s_t$, i.e. $\mathbf{h}'_{s} = f_{MLP}(\mathbf{h}_{s})$. To ensure training stability, we stop the gradient flow through $\mathbf{h}'_{s}$ for the actor network and only use the more complex critic networks to optimize $\mathbf{h}'_{s}$, as we find that sharing the feature extraction module and the $f_{MLP}$ network causes interference between the policy and critic gradients.

\subsection{Multi-Group State Decomposition Module}
Since live stream content is not equally appealing to all users of the short video app \textit{Kwai}, it is crucial to inject live streams selectively into the short video feed. Otherwise, excessive live stream exposures may disrupt a user's interest in short videos, leading to a decrease in overall app usage duration. In practice, user interaction data for live streams are sparser and noisier compared to that for short videos, making it challenging for RL algorithms to learn an optimal live stream allocation policy for each individual user. Directly learning a policy from sparse and noisy user feedback is often infeasible, since it will inadvertently inject excessive live streams in the short video feed and highly affected the user engagement. 

We find that prior information on diverse user groups is essential to improve the accuracy of the decision making of RL models in our scenario, which provides additional user preference information to the policy. A natural approach to distinguish between different user behaviors is to partition users into distinct groups based on their historical activity. Specifically, we categorize users into $K$ disjoint groups according to their activity level of the live stream, which is determined by the cumulative watch time of the live stream during the past 3 weeks. Users with a higher historical watch time are assigned to groups with higher activity levels.

% Formally, we present the following theorem, which demonstrates that the proposed multi-group state decomposition can be viewed as a type of \textit{variance reduction} technique for real-world RL applications.

% \begin{theorem}
% Let state space $\mathcal{S} = \bigcup_{i = 1}^{K} \mathcal{S}_i, \forall i \neq j \; \mathcal{S}_i \cap \mathcal{S}_j = \varnothing$, state $s \in \mathcal{S}$; $r$ be the real reward of state $s $ and action $a$, $\hat{r}$ be the predicted value of $r$ over $\mathcal{S}$, $\hat{r}_{i}$ be the predicted reward of $r$ over $\mathcal{S}_i$; then a neural network $f$ with MG-SD module has a lower prediction variance than the neural network $f'$ without MG-SD module.
% \label{theorem:1}
% \end{theorem}

Empirically, we show the effectiveness of this module in enhancing the stability of the RL model in subsequent experiments. We find that SL-MGAC without the MG-SD module tends to be more unstable during training, with a larger variance in its Q-values. Furthermore, the MG-SD module is flexible and transferable, allowing the user-group partitioning strategy to be easily adapted to various scenarios in advertising and E-commerce.

\subsection{Supervised Learning-enhanced Actor Critic}
To alleviate the impact of drastic changes in the data distribution that could lead to divergence or instability of RL models, we propose a supervised learning-enhanced actor critic framework to prevent critic networks from being trapped in model collapse due to large cumulative errors in the critic learning process.

\subsubsection{Layer Normalization}
A recent work on RL divergence and instability \cite{yue2024understanding} shows that offline critic learning exhibits a \textit{self-excitation} pattern. Specifically, iteration of the Q value can inadvertently cause the target Q value $Q(s_{t+1}, a^{*})$ to increase even more than the increment of $Q(s_t, a_t)$, amplifying the TD error and trapping the critic learning process in a positive feedback loop until the model collapse. Therefore, normalization techniques, such as \textit{Layer Normalization} \cite{ba2016layer} can be utilized to alleviate divergence and instability problems. 

From the proof in Appendix D of \cite{yue2024understanding}, we know that for any input $\mathbf{x}$ and any direction $\mathbf{v}$, if a network applies Layer Normalization to the input, then we have $k_{\rm NTK}(\mathbf{x}, \mathbf{x} + \lambda \mathbf{v}) \leq C$, where $\lambda > 0$, $C$ is a constant, and $k_{\rm NTK}$ is the \textit{Neural Tangent Kernel} (NTK) \cite{jacot2018neural}. This theoretically indicates that a network with Layer Normalization is less sensitive to input variations and can maintain stable gradients despite perturbations during model training. Hence, we apply Layer Normalization to the inputs of both the actor and critic networks, as shown in Fig. \ref{fig:3}.

% In practice, we also attempt another normalization technique, such as Logit Normalization \cite{wei2022mitigating}, which applies L2-normalization to the logit output of the actor network. However, Logit Normalization may fail when the network gradient of SL-MGAC is large, leading to the phenomenon of \textit{policy deterioration}, where $\forall t, \; \pi(a_t  = 1|s_t) = 0$ or $\pi(a_t  = 1|s_t) = 1$. 

\subsubsection{Critic Network}
Let $(s_t, a_t, r_t, s_{t+1}, a_{t + 1}, r_{t + 1}) \in \mathcal{D}$ be a training sample from our real-time dataset $\mathcal{D}$. To address the \textit{maximization bias} problem \cite{van2016deep}, we employ four critic networks: two current Q-networks $Q_{\phi_1}$, $Q_{\phi_2}$ and two corresponding target Q-networks $Q'_{\phi_1}$, $Q'_{\phi_2}$ in Clipped Double Q-Learning \cite{fujimoto2018addressing}. The critic learning objective is as follows:
\begin{equation}
\begin{aligned}
\mathcal{L}_{Critic} = \sum_{i=1}^{2} \mathbb{E}_{(s, a) \in \mathcal{D}}&\left[Q_{\phi_i}(s_t, a_t) - Q_{label}(s_{t+1})\right]^2 \\
Q_{label}(s_{t+1}) &= r(s_t, a_t) + \gamma \max_{a_{t+1}} \hat{Q}'(s_{t+1}, a_{t+1})
\end{aligned}
\label{eq:9}
\end{equation}
where $\hat{Q}'(s_{t+1}, a_{t+1}) = \min _{i=1,2} Q'_{\phi_i}\left(s_{t+1}, a_{t+1}\right)$. We use the Huber loss \cite{hastie2009elements} to optimize the above objective. 

In practice, $Q_{label}(s_{t+1})$ may be dominated by $\hat{Q}'(s_{t+1}, a_{t+1})$ during critic learning, which can affect the performance of critic networks. Specifically, due to instability and divergence issues, target networks often predict inaccurate $\hat{Q}'(s_{t+1}, a_{t+1})$ values, which is much larger than the $r(s_t, a_t)$ term. Then the term $r(s_t, a_t)$ could not provide any information to guide the learning of the critic. Therefore, we seamlessly introduce supervised learning (e.g. multi-task learning \cite{zhang2021survey}) into the critic learning procedure. Specifically, we divide the critic network $Q_{\phi_i}$ (or $Q'_{\phi_i}$) into two components: a Reward Prediction Network (RPN) and a Q Residual Network (QRN), as follows:

\begin{equation}
\begin{aligned}
Q_{\phi_i}(s_t, a_t) &= R_{\theta_i}(s_t, a_t) + \gamma T_{\xi_i}(s_t, a_t), \quad i=1,2 \\
Q'_{\phi_i}(s_{t+1}, a_{t+1}) &= R'_{\theta_i}(s_{t+1}, a_{t+1}) + \gamma T'_{\xi_i}(s_{t+1}, a_{t+1}), \quad i=1,2
\end{aligned}
\label{eq:10}
\end{equation}
where $R_{\theta_i}$ and $T_{\xi_i}$ are RPN and QRN, $R'_{\theta_i}$ and $T'_{\xi_i}$ are target RPN and QRN, respectively. We use separate MLPs to model the RPN and QRN.

Since the distribution of live stream (short video) watch time changes drastically over time, the time gain reward $r$ in Eq. \ref{eq:8} becomes difficult to learn. In this work, we propose a \textit{distribution discretization} method to improve reward learning. 

Formally, we divide the live stream and short video watch time distribution into $N_{l}$ and $N_{v}$ non-overlapping bins, where each bin represents an interval of live stream (short video) watch time. Let $y$ denote the actual live stream or short video watch time, which falls in the time bin $[y_{st}, y_{end}]$. The proportion of $y$ within this bin is $\delta = \frac{y - y_{st}}{y_{end} - y_{st}} \in [0, 1]$. Then, we have a \textit{linear reconstruction layer} to reconstruct $y$ from $\delta$:
\begin{equation}
\begin{aligned}
y &= y_{st} + \delta \cdot (y_{end} - y_{st}) \\
\end{aligned}
\label{eq:11}
\end{equation}
which resembles the structure of linear reward shifting \cite{sun2022exploit}.

Let $o(t)_{l} \in \mathbb{R}^{N_{l} + 1}$, $o(t)_{v} \in \mathbb{R}^{N_{v} + 1}$ be one-hot vectors representing the time bins in which the real live stream (short video) watch time of sample $(s_t, a_t, r_t)$ falls. Note that a separate bin is set for the case $a_t = 0$. Then, the RPN $R_{\theta_i}$ can be modeled by multi-task neural networks:
\begin{equation}
\begin{aligned}
R_{\theta_i}(s_t, a_t) &= sigmoid\left(F_{\Gamma_i}(s_t, a_t) - G_{\Theta_i}(s_t, a_t)\right) \\
F_{\Gamma_i}(s_t, a_t) &= {o(t)_{l}}^{T} \left(y_{l}^{st} + f_{\Gamma_i}(s_t, a_t) \cdot \left(y_{l}^{end} - y_{l}^{st}\right)\right) \\
G_{\Theta_i}(s_t, a_t) &= {o(t)_{v}}^{T} \left(y_{v}^{st} + g_{\Theta_i}(s_t, a_t) \cdot \left(y_{v}^{end} - y_{v}^{st}\right)\right) \\
\end{aligned}
\label{eq:12}
\end{equation}
where $y_{l}^{st}, y_{l}^{end} \in \mathbb{R}^{N_{l} + 1}$ predefined left (right) boundary vectors for the live stream watch time bin, $y_{v}^{st}, y_{v}^{end} \in \mathbb{R}^{N_{v} + 1}$ are predefined left (right) boundary vectors for the short video watch time bin. Note that we introduce the posterior one-hot vectors $o(t)_{l}$ and $o(t)_{v}$ in Eq. \ref{eq:12} will not affect the calculation of $Q'_{\phi_i}(s_{t+1}, \cdot)$, because we have already reserved $r_{t+1}$ in the dataset $\mathcal{D}$. Therefore, $o(t+1)_{l}$ and $o(t+1)_{v}$ can be easily obtained from $r_{t+1}$ to compute $R'_{\theta_i}(s_{t+1}, \cdot)$.

As shown in Eq. \ref{eq:12}, we employ two separate MLPs, $f_{\Gamma_i}(s_t, a_t)$ and $g_{\Theta_i}(s_t, a_t)$, to predict the proportions within a time bin. We then reconstruct the predicted live streaming and short video watch times, $F_{\Gamma_i}(s_t, a_t)$ and $G_{\Theta_i}(s_t, a_t)$, respectively. Finally, we obtain the predicted normalized reward, $R_{\theta_i}(s_t, a_t)$. To optimize the predicted watch-time proportions, we use Huber loss as follows:

\begin{equation}
\begin{aligned}
\mathcal{L}_{SL} = \sum_{i=1}^{2} Huber\_Loss(\delta_{l}, F'_{\Gamma_i}) + Huber\_Loss(\delta_{v}, G'_{\Theta_i})\\
\end{aligned}
\label{eq:13}
\end{equation}
where $\delta_{l}$, $\delta_{v}$ are the actual proportion labels for live stream and short video watch times, respectively.

The reason for using the Huber loss on the time ratio $\delta$ instead of the actual watch time $y$ is that the time ratio is within the range $[0, 1]$, resulting in smaller gradients in the neural network. Furthermore, the variance of $\delta$ is much smaller than that of the original watch time $y$. Consequently, the outputs of the learned reward networks ($f_{\Gamma_i}$ and $g_{\Theta_i}$) will exhibit smaller variances. Hence, reward learning with the discretization of watch time distributions can be viewed as a novel \textit{variance reduction} technique.

\subsubsection{Actor Network}
\label{sbsc:an}
We also incorporate the multi-group state decomposition module into the actor network. The loss function of the actor network is shown below:

\begin{equation}
\begin{aligned}
\mathcal{L}_{Actor} &= \mathbb{E}_{(s, a) \in \mathcal{D}} \left[-\hat{Q}(s_t, a_t) \log p(s_t, a_t)\right] \\
\label{eq:14}
\end{aligned}
\end{equation}
where $\hat{Q}(s_t, a_t) = \min_{i=1,2} Q_{\phi_i}(s_t, a_t)$, and $p(s_t, a_t)$ is the action probability output of the actor network. 

Moreover, we observe that high values of $\hat{Q}(s_t, a_t)$ can cause instability in actor training, which can ultimately lead to \textit{policy deterioration}, where $\forall t, \; \pi(a_t  = 1|s_t) > \pi(a_t  = 0|s_t)$ or $\pi(a_t  = 1|s_t) < \pi(a_t  = 0|s_t)$. To address this, we apply softmax normalization to $\hat{Q}(s_t, a_t)$ to obtain $\hat{Q}_{norm} = softmax(\hat{Q}(s_t, a_t))$, and propose a modified actor loss function:

\begin{equation}
\begin{aligned}
\mathcal{L}_{Actor} &= \mathbb{E}_{(s, a) \in \mathcal{D}} \left[-\hat{Q}_{norm}(s_t, a_t) \log p(s_t, a_t)\right] \\
\label{eq:15}
\end{aligned}
\end{equation}

A similar loss function can be found in AWAC \cite{nair2020awac}, and its theoretical results demonstrate that the above loss is equivalent to an implicitly constrained RL problem:

\begin{equation}
\begin{aligned}
\pi_{k + 1} &= \arg \max_{\pi \in \Pi} \mathbb{E}_{\mathbf{a} \sim \pi(\cdot |\mathbf{s})}\left[\hat{Q}^{\pi_{k}}(s_t, a_t)\right] \\
{\rm s.t.} \; & D_{\rm KL}\left(\pi(\cdot | \mathbf{s}) \| \pi_{o}(\cdot | \mathbf{s})\right) \leq \epsilon \\
\label{eq:16}
\end{aligned}
\end{equation}
where $D_{\rm KL}$ is the Kullback-Leibler (KL) divergence, and $\pi_{o}$ is the behavior policy derived from the dataset $\mathcal{D}$.

We note that the actor loss in Eq. \ref{eq:15} is a standard cross-entropy loss function. From the perspective of knowledge distillation \cite{hinton2015distilling}, the actor distills policy knowledge from more complex critic networks. The teacher critic networks guide the more lightweight student actor network to adjust and converge to an optimal policy. In practice, we only need to deploy the actor network in the online live stream recommendation system, which significantly reduces computational complexity and improves real-time response speed.

Overall, the final loss function of our proposed SL-MGAC algorithm is as follows:
\begin{equation}
\begin{aligned}
\mathcal{L} = \mathcal{L}_{Actor} + \mathcal{L}_{Critic} + \mathcal{L}_{SL} \\
\end{aligned}
\label{eq:17}
\end{equation}

\subsection{Online Exploration and Deployment}

For online exploration, we adopt the commonly used $\epsilon$-greedy \cite{watkins1989learning} strategy:

\begin{equation}
\pi_{online}(s_t) =  \begin{cases}
{\rm random\; action\; from\;} \mathcal{A}(s_t), \quad &if\; \psi < \epsilon \\
\arg\max_{a \in \mathcal{A}(s_t)} \pi(a_t | s_t), &{\rm otherwise}
\end{cases}
\label{eq:18}
\end{equation}
where $\psi$ is a random number, and $\epsilon$ is maximal exploration probability.

\begin{figure}[htbp]
\centering
\includegraphics[width=0.48\textwidth]{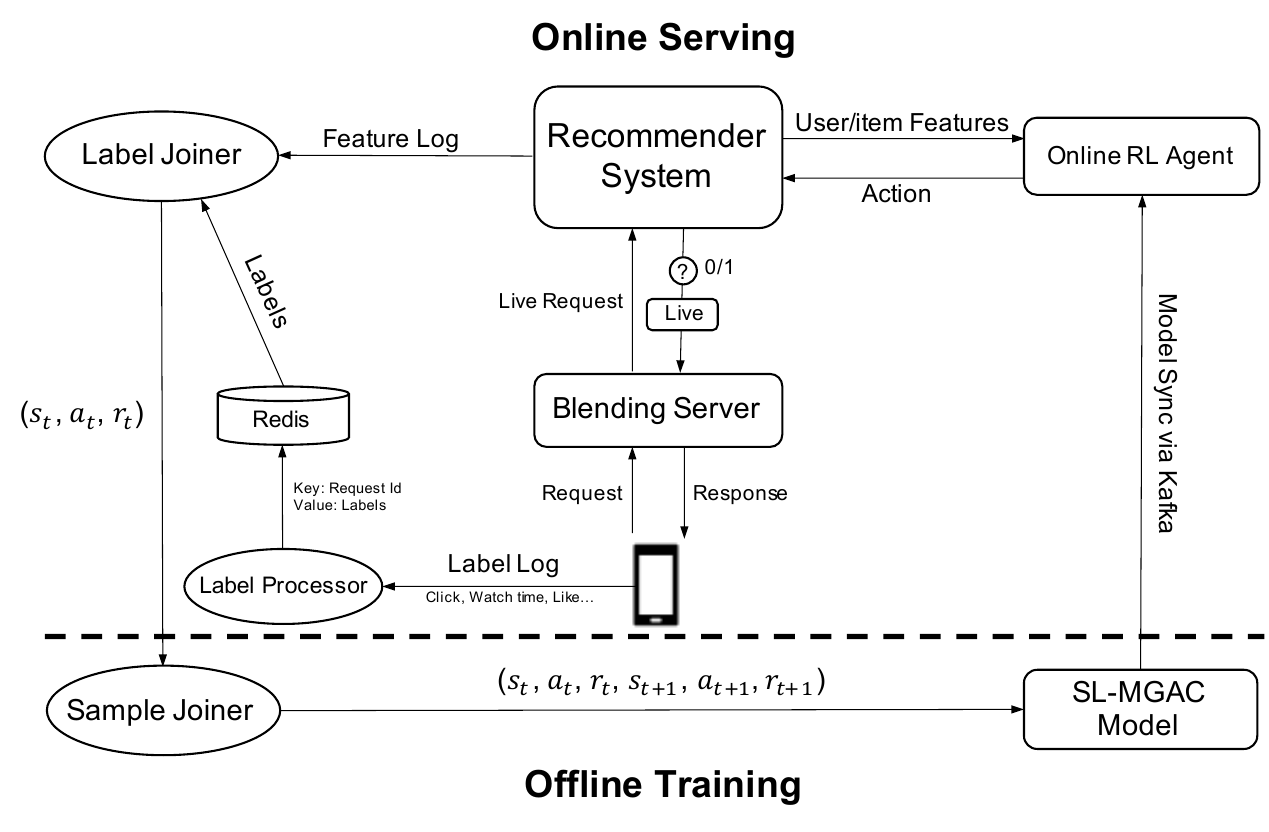}
\caption{System Architecture of the SL-MGAC algorithm. }
\label{fig:4}
\end{figure}

We implemented the SL-MGAC algorithm in our recommendation system, and the overall architecture of the system is shown in Fig. \ref{fig:4}. The online RL agent collects real-time user interaction logs, while the offline model trainer optimizes the SL-MGAC model in an off-policy manner using streaming training data. Moreover, the offline trainer sends the latest model parameters to update the online deployed actor network in real-time.

Most importantly, we focus on reducing computational costs for successful deployment in two key aspects. First, for real-time training, we avoid using user-item cross features, which are commonly employed in industrial ranking models, thereby reducing the sparse embedding parameters. Additionally, the network parameters of the Actor part in SL-MGAC are much smaller than those of the Critic, with the complex Critic acting as a teacher to guide the Actor's learning, as explained in Section \ref{sbsc:an}. As a result, the total parameters of SL-MGAC are under 100 million in our recommendation system, significantly smaller than the parameter scale of typical ranking models, which usually exceed 1 billion parameters.% For offline training, we only use two machines with 100 CPU cores.

For fast online inference, the smaller Actor network enables fast inference, processing each request in under 20ms, even with over 10,000 requests per second during peak times. Furthermore, since the complex Critic is not used during online inference, this further reduces computational costs, contributing to faster response time in the online environment.

\section{Experiments}
We perform both offline evaluation on a real-world dataset collected from our recommendation system and online A/B test experiments with SL-MGAC and the baselines.

\subsection{Dataset}
Due to the lack of publicly available datasets for recommendation decisions involving live streaming allocations in short video feeds, we collect an offline dataset from our recommendation system with 100,000 users through random sampling over a full day for offline evaluation, ensuring that the proportions of different user groups are similar to those in real online data. In total, we have over 1,800,000 samples, which are divided into training and test sets with a 4:1 ratio. In addition, the online environment contains more than 100 million app users for AB-Tests.

\subsection{Compared Methods}
\subsubsection{Baselines}
We compare our approach with existing non-reinforcement learning and reinforcement learning methods. 
\begin{itemize}
\item \textbf{Learning to Rank (L2R)} \cite{chen2020deeprank}. A supervised learning framework that predicts the reward for each action. The action with the maximum reward is selected as the agent's action.
\item \textbf{DQN} \cite{mnih2015human}. A deep neural network algorithm for Q-learning that introduces the technique of updating the target network.
\item \textbf{BCQ} \cite{fujimoto2019off}. A widely used offline reinforcement learning algorithm that adds action restrictions and encourages the agent to behave similarly to the behavior policy.
\item \textbf{SAC} \cite{haarnoja2018soft, christodoulou1910soft}. A classic off-policy reinforcement learning method that maximizes the trade-off between cumulative reward and policy entropy. We use the discrete version of SAC in later experiments.
\item \textbf{TD3} \cite{fujimoto2018addressing}. A modified version of DDPG \cite{lillicrap2015continuous} addresses function approximation errors using three techniques: Clipped Double Q-Learning, Delayed Policy Updates, and Target Policy Smoothing.
\item \textbf{TD3-BC} \cite{fujimoto2021minimalist}. An offline reinforcement learning variant of the TD3 algorithm, with behavior cloning (BC) regularization to constrain policy learning.
\item \textbf{IQL} \cite{kostrikov2021offline}. An offline reinforcement learning algorithm that leverages expectile regression in Q-Learning.
\item \textbf{RLUR} \cite{cai2023reinforcing}. An RL model that aims to optimize the long-term user retention.
\end{itemize}
Note that in our optimal live stream allocation control problem, the action is discrete. Therefore, we introduce the Straight-Through Gumbel Softmax \cite{jang2016categorical} technique in TD3 and TD3-BC.

\subsubsection{Variations of our model}
We also compare the SL-MGAC algorithm with several variants in an ablation study to illustrate the effectiveness of multi-group state decomposition, supervised learning for critic learning, distribution discretization (DD), and other techniques.

\begin{itemize}
\item \textbf{SL-MGAC (w/o MG)}: An SL-MGAC variant without the Multi-Group State Decomposition module.
\item \textbf{SL-MGAC (w/o MG \& DD)}: An SL-MGAC variant without the Multi-Group State Decomposition module and the distribution discretization technique in reward learning.
\item \textbf{SL-MGAC (w/o MG \& DD \& SL)}: An SL-MGAC variant without the Multi-Group State Decomposition module, the distribution discretization technique in reward learning, and the supervised learning procedure.
\item \textbf{SL-MGAC (w/o LN)}: An SL-MGAC variant without the Layer Normalization technique.
\item \textbf{SL-MGAC (w/o SG)}: An SL-MGAC variant without the Stop Gradient technique for hidden state representation in the actor network.
\item \textbf{SL-MGAC (w/o Q-norm)}: An SL-MGAC variant without Q-value normalization in actor loss.
\item \textbf{SL-MGAC-0}: An SL-MGAC variant with the discount factor $\gamma = 0$.
\item \textbf{SL-MGAC-sep}: An SL-MGAC variant with separate optimization of the actor network. Hence, $\hat{Q}_{norm}$ in Eq. \ref{eq:15} is detached from the computation graph.
\item \textbf{SL-MGAC-vanilla}: An SL-MGAC variant with the vanilla Q label in the loss of the critic, as shown below, where $p(s_{t+1}, \cdot)$ is the probability of action of the output of $s_{t+1}$ by the actor network.
\begin{equation}
\begin{aligned}
Q_{label}(s_{t+1}) &= r(s_t, a_t) + \gamma \sum_{a_{t+1}} p(s_{t+1}, a_{t+1})\hat{Q}(s_{t+1}, a_{t+1})
\end{aligned}
\label{eq:19}
\end{equation}
\end{itemize}

\subsection{Implementation Details}
To ensure fairness among all compared methods, we use a consistent feature extraction module with a 5000-size embedding layer. In addition to the embedding layer, the feature extraction module includes a 2-layer MLP with hidden sizes of [256, 128]. The batch size and number of epochs for all methods are 2048 and 500, respectively. The learning rate for the embedding layer is set to 1e-5, while the learning rate for other hidden parameters is 1e-3. The discount factor $\gamma$ is set to 0.9 for all methods and $\lambda$ in Eq \ref{eq:8} is 0.1. Detailed network architecture and parameter settings for SL-MGAC are provided in Table \ref{table:4} of the Appendix \ref{Appdenx: B}. And the code of SL-MGAC method is available at \href{https://github.com/frankg1/SL-MGAC-torch}{https://github.com/frankg1/SL-MGAC-torch}.

\subsection{Offline Policy Evaluation}
We follow the approach in \cite{xue2023prefrec} and adopt a commonly used offline policy evaluation method, namely Normalized Capped Importance Sampling (NCIS) \cite{swaminathan2015self}, to evaluate performance. The evaluation metric is the cumulative reward across all trajectories of test users.

\begin{table}[htbp]
\centering
\begin{tabular}{cccc}
\hline
Methods & Cumulative Reward \\
\hline
L2R \cite{chen2020deeprank} & 413.49 \\
DQN \cite{mnih2015human}   & 413.18  \\
BCQ \cite{fujimoto2019off}   & 416.69  \\
SAC \cite{christodoulou1910soft}   & 435.41 \\
TD3 \cite{fujimoto2018addressing} & 432.08  \\
TD3-BC \cite{fujimoto2021minimalist} & 430.23  \\
IQL \cite{kostrikov2021offline} & 432.03  \\
RLUR \cite{cai2023reinforcing} & \textbf{443.01} \\
\hline
SL-MGAC-sep & 435.19 \\
SL-MGAC-vanilla & 450.21 \\
\textbf{SL-MGAC} & \textbf{458.49} \\
\hline
\end{tabular}
\caption{Overall offline performance of compared methods.}
\label{table:1}
\end{table}

The results of the offline policy evaluation are shown in Table \ref{table:1}. Compared to the supervised learning method L2R, most RL methods achieve higher cumulative rewards, except for DQN. SL-MGAC significantly outperforms all other methods, demonstrating that SL-MGAC can achieve higher long-term rewards in the complex live stream \& short video mixed recommendation system with drastically changing data distributions. Furthermore, SL-MGAC outperforms SL-MGAC-vanilla, suggesting that incorporating the actor network in the Q label calculation of Eq. \ref{eq:19} may interfere with the learning of the critic to some extent, slightly affecting the performance of the SL-MGAC-vanilla. % Furthermore, SL-MGAC converges faster than SL-MGAC-sep during training, as shown in Fig. \ref{fig:7} in Appendix \ref{Appdenx: C}, and achieves a higher cumulative reward in offline evaluation. This may be due to information loss caused by detaching $\hat{Q}_{norm}$, which can lead to slower convergence for SL-MGAC-sep.

\subsection{Ablation Study}
We compare the offline performance of different SL-MGAC variants. As shown in Table \ref{table:2}, SL-MGAC outperforms all other variants on our offline dataset, demonstrating the effectiveness of the proposed multi-group state decomposition module, supervised learning procedure, and other techniques. Moreover, we find that SL-MGAC (w/o Q-norm) achieves the lowest reward, indicating that the Q-normalization technique in the actor loss enhances the model's convergence and improves its performance.

\begin{table}[htbp]
\centering
\begin{tabular}{cccc}
\hline
Methods & Cumulative Reward \\
\hline
SL-MGAC (w/o MG) &  454.62 \\
SL-MGAC (w/o MG \& DD) & 452.59 \\
SL-MGAC (w/o MG \& DD \& SL) & 449.13 \\
\hline
SL-MGAC (w/o LN) & 453.61 \\
SL-MGAC (w/o SG) & 457.26 \\
SL-MGAC (w/o Q-norm) & 392.12 \\
\hline
\textbf{SL-MGAC} & \textbf{458.49} \\
\hline
\end{tabular}
\caption{Offline performance of SL-MGAC variants.}
\label{table:2}
\end{table}

\begin{figure}[htbp]
\centering
\includegraphics[width=0.48\textwidth]{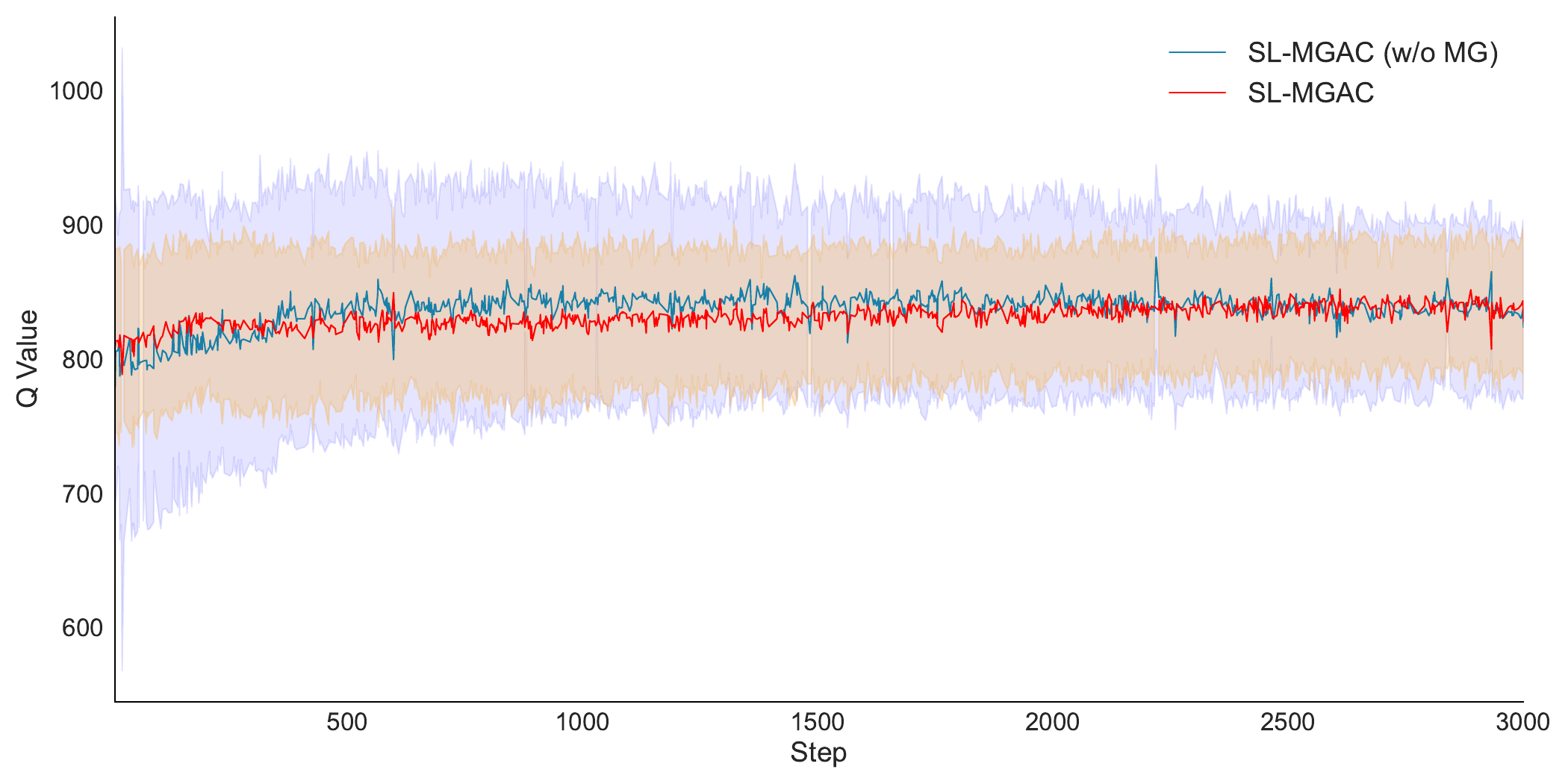}
\caption{The Q value curves between SL-MGAC and SL-MGAC (w/o MG) over 10 rounds of training. The lines correspond to the means of Q-value and the shaded areas correspond to the standard deviations (std).}
\label{fig:10}
\end{figure}

Next, we compare the training processes of SL-MGAC and SL-MGAC (w/o MG), as shown in Fig. \ref{fig:10}. Note that the shaded area in yellow corresponds to the Q value std range of the SL-MGAC model, while the shaded area in purple corresponds to the Q value std range of the SL-MGAC model(w/o MG). We observe that the Q-value curve of SL-MGAC is more stable than that of SL-MGAC (w/o MG), and the Q-value variance of SL-MGAC is much smaller. This demonstrates the effectiveness and variance reduction effect of the MG-SD module.

\subsection{Parameter Sensitivity}
\begin{figure}[htbp]
\centering
\includegraphics[width=0.48\textwidth]{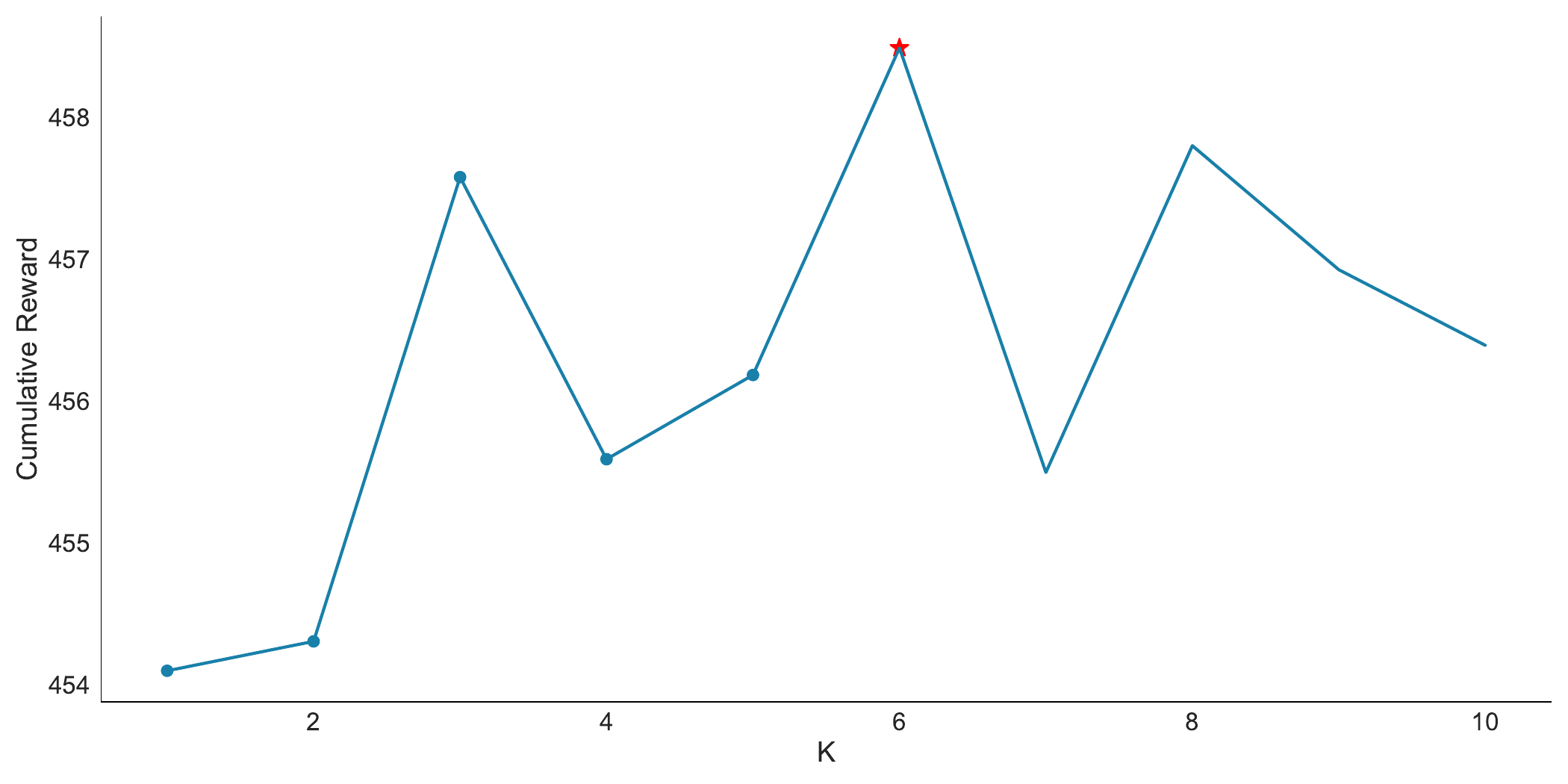}
\caption{Performance of different numbers of user group $K$.}
\label{fig:5}
\end{figure}

We analyze the impact of the number of user groups $K$ on the performance of SL-MGAC. As shown in Fig. \ref{fig:5}, the proposed SL-MGAC algorithm with $K = 6$ achieves the highest cumulative reward, demonstrating that increasing the number of user groups within a certain range improves model performance.

\begin{table*}[htbp]
\centering
\begin{tabular}{ccccccc}
\hline
Methods & live stream DAU & live stream Watch Time & Video Watch Time & App Usage Duration & User Retention \\
\hline
L2R & +0.321\% & -0.558\% & -0.238\% & -0.178\% & -0.154\% \\
Dummy & \textbf{+51.3\%}  & \textbf{+34.6\%}  & -2.216\% & -1.062\% & -0.439\% \\
SL-MGAC-0 & +3.928\% & -2.934\%  &-0.153\% &-0.117\% & -0.102\% \\
SL-MGAC & +2.616\% & +7.431\% &\textbf{+0.197\%} & \textbf{+0.121\%} & \textbf{+0.086\%} \\
\hline
\end{tabular}
\caption{Online A/B Test Performance of compared methods.}
\label{table:3}
\end{table*}

\subsection{Online A/B Test Experiment}
The online A/B Test is conducted over a 5-day period in 2024 September and we randomly choose 20\% users (over 20 million) in Kwai app as the experimental group. Then we compare the improvements of different methods in terms of daily active live stream users (DAU), live stream watch time, video watch time, app usage duration, and user retention relative to the baseline. The baseline uses the SAC framework for live stream allocation. The results of the online A/B test are shown in Table \ref{table:3}. 

Note that the Dummy method injects a live stream for each request. We can see that SL-MGAC achieves a significant improvement in live stream watch time while also increasing the app usage duration and user retention than other methods. This shows that SL-MGAC is more effective in maximizing long-term live stream rewards while avoiding the negative impact on app usage duration and user retention. Although SL-MGAC-0 shows a greater improvement in live stream DAU compared to SL-MGAC, it tends to be more greedy in injecting live streams, which may affect the long-term user experience for most users. Moreover, although the dummy method can greatly improve DAU and live stream viewing time, it significantly affects app usage duration and user retention.

% Since the scale of live stream supply and user interaction behaviors change drastically over time, more streamers tend to broadcast live during peak hours. As shown in Fig. \ref{fig:8} in Appendix \ref{Appdenx: C}, SL-MGAC can adaptively adjust the live stream injection strategy to increase live stream exposure during peak hours, while reducing exposure during low-traffic periods.

\subsection{Online Model Stability}
\begin{figure}[h!]
\centering
\includegraphics[width=0.48\textwidth]{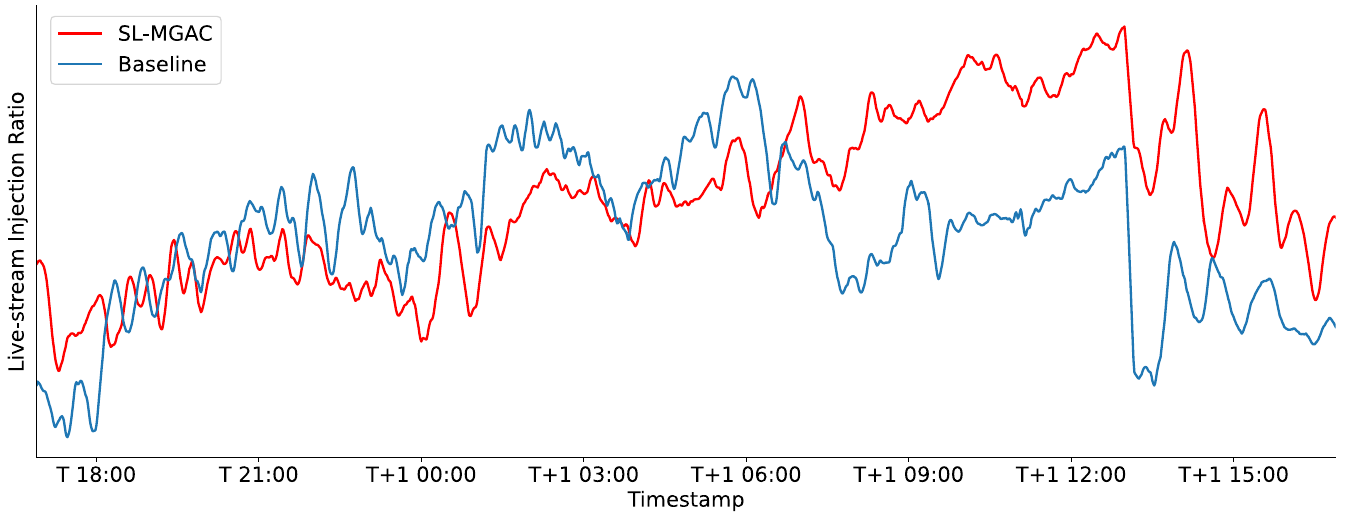}
\caption{The trend curve of online live stream allocation ratio during a whole day. }
\label{fig:6}
\end{figure}

We evaluate the online model stability between the SAC-based baseline and SL-MGAC. The live-stream injection ratio is chosen as the evaluation metric, calculated by dividing the number of live-stream injection requests by the total number of requests within a given timestamp. As shown in Fig. \ref{fig:6}, the live-stream allocation ratio for both models fluctuates significantly throughout the day, from high-traffic hours (such as 6:00–11:00) to low-traffic hours (14:00–18:00). This variation is due to changes in the overall user scale during these periods, which in turn affect the data distribution. However, SL-MGAC exhibits greater stability than the baseline, with a smaller amplitude. Note that the live stream allocation ratio refers to the proportion of requests with action $a = 1$.

We calculate the amplitudes of the live stream allocation ratio within sliding time windows of 20 minutes. The amplitude density is shown in Fig. \ref{fig:9} in Appendix \ref{Appdenx: C}. The results indicate that SL-MGAC has a smaller mean amplitude and fewer outliers. In contrast, the baseline model exhibits larger amplitude outliers, as highlighted in the red dashed box, demonstrating the superior online model stability of SL-MGAC.

%\section{Discussion}
%The intrinsic instability and divergence of RL limits its application in real-world industrial recommendation systems. The proposed SL-MGAC method improves the robustness and stability of the RL model and achieves significant offline and online performance in the live stream and short video mixed recommendation scenario. However, SL-MGAC only learns from data from adjacent states $(s_t, a_t, r_t, s_{t+1}, a_{t+1}, r_{t+1})$, which limits its ability to learn information in much longer sequences, such as the daily-level user preference, rather than session-level user preference, since the capture of daily user preference information is very helpful for improving user engagement and user retention. 

%Furthermore, we emphasize that more sophisticated methods can also be employed to divide users into multiple groups and provide prior information about user preference, such as traditional clustering methods, e.g. k-means, neural clustering of VQ-VAE \cite{van2017neural}, online clustering of SwAV \cite{caron2020unsupervised} by the Sinkhorn-Knopp algorithm \cite{cuturi2013sinkhorn}, etc. We find that the simple MG-SD module achieves significant performance and fast real-time response speed in practice; we will lead these complicated group partitioning methods in our further research.

\section{Related Work}
\subsection{RL in Recommendation Systems}
Reinforcement learning (RL) aims to optimize cumulative long-term rewards over time, which has attracted significant attention in the research of recommendation systems in recent years \cite{afsar2022reinforcement}. Methods such as SLATEQ \cite{ie2019slateq}, GeMS \cite{deffayet2023generative}, and HAC \cite{liu2023exploration} use RL to recommend complete item lists, where the number of candidate items can be large. BatchRL-MTF \cite{zhang2022multi}, RLUR \cite{cai2023reinforcing}, and UNEX-RL \cite{zhang2024unex} leverage RL to model the multi-rank score aggregation process, optimizing the weights for score aggregation. The work in \cite{miao2021sequential} explores the use of off-policy RL for personalized search ranking in multiple sessions. CrossDQN \cite{liao2022cross} introduces an RL-based approach to ad allocation in a feed, aiming to maximize revenue while improving the user experience. Furthermore, traditional RL methods such as DQN \cite{mnih2013playing, mnih2015human}, Double DQN \cite{van2016deep}, SAC \cite{haarnoja2018soft}, DDPG \cite{lillicrap2015continuous}, and TD3 \cite{fujimoto2018addressing} serve as backbones in real-world RL applications for recommendation systems.

A similar approach, called self-supervised actor-critic \cite{xin2020self}, combines supervised learning with critic learning. Their supervised learning task focuses on predicting the next item recommendation probability, whereas our approach introduces supervised learning to restrict critic learning.

However, the aforementioned RL methods may encounter instability issues and could fail when faced with highly noisy or drastically changing data distributions. The proposed SL-MGAC method is more robust and has been successfully deployed in the "risky" final stage of a real-world industrial RS. We believe this work provides valuable insights not only into the model framework design for RL applications in industrial RS, but also offers practical experience in deploying robust and successful RL systems for other industrial companies.

\subsection{RL divergence and instability}
Recently, there has been a growing literature that focuses on RL divergence and instability. \cite{van2016deep} introduces Double DQN to mitigate the \textit{maximization bias} problem in DQN. \cite{fujimoto2018addressing} proposes the Clipped Double Q-Learning technique in TD3 to reduce the overestimation of Q values. \cite{mao2018variance} introduces a bias-free, input-dependent baseline for the policy gradient algorithm \cite{sutton1999policy} to reduce variance and improve the stability of training. \cite{dohare2023overcoming} investigates policy collapse in a 2-state MDP and finds that L2 regularization and the non-stationary Adam optimizer \cite{kingma2014adam} are both effective in alleviating RL instability. \cite{yue2024understanding}  theoretically analyzes the causes of RL divergence and applies Layer Normalization to mitigate RL divergence and instability. 

\section{Conclusion}
In the challenging context of a live stream \& short video mixed recommendation system, we propose a novel Supervised Learning-enhanced Multi-Group Actor-Critic algorithm (SL-MGAC) to optimize request-level live stream allocation policies under the platform-level constraints on app usage duration and user retention. Specifically, we introduce a multi-group state decomposition module to reduce prediction variance and seamlessly integrate multi-task supervised reward learning with traditional critic learning to constrain Q-value estimation. Compared with existing RL methods for recommendation systems, our approach not only focuses on improving online performance, but also improves the stability and robustness of the RL model. In practice, we minimize the risk of \textit{policy deterioration} or \textit{model collapse}, thereby enabling the proposed SL-MGAC method to be successfully implemented in large-scale industrial recommendation systems. Currently, we are trying to apply the SL-MGAC methods to other similar recommendation scenarios, including E-commerce and advertising. We also aim to continuously improve the performance of SL-MGAC in further research. Learning live stream policy from a long sequence of user behaviors $(s_t, a_t, r_t)$ and exploring more refined and complex user state representations would lead to our future work. 

\section{Acknowledgment}
The authors acknowledge Wei Bai, Xiaoshuang Chen and anonymous reviewers for proposing detailed modification advice to help us improve the quality of this manuscript. 

\bibliographystyle{ACM-Reference-Format}
\balance
\bibliography{reference}

\newpage
\appendix
\balance
\onecolumn
\section{Tables}
\label{Appdenx: B}

\begin{table}[h!]
\centering
\begin{tabular}{cccc}
\hline
Hyper-parameter & Value \\
\hline
MGN in Actor & [128, 63, 31, 2] \\
MGN in RPN  & [128, 64, 32, 8] \\
MGN in QRN   & [128, 64, 32, 2]  \\
Live-Time DDB & [0s, 6s, 15s, 30s, 60s, 100s, 600s, 1200s] \\
Video-Time DDB & [0s, 3s, 10s, 25s, 50s, 100s, 600s, 1200s] \\
Optimizer & Adam \\
User Group Number $K$ & 6 \\
Online Exploration $\epsilon$ & 0.2 \\
\hline
\end{tabular}
\caption{Hyper-parameters of SL-MGAC.}
\label{table:4}
\end{table}

We denote MGN as the multi-group network, RPN as the Reward Prediction Network, QRN as Q Residual Networks, and DDB as Distribution Discretization Bins. Both live and video watch time distribution have 7 time bins. We leave a separate time bin for the $a = 1$ case, and hence the output layer dimension of RPN is 8.

\section{Figures}
\label{Appdenx: C}

\begin{figure}[htbp]
\centering
\includegraphics[width=0.6\textwidth]{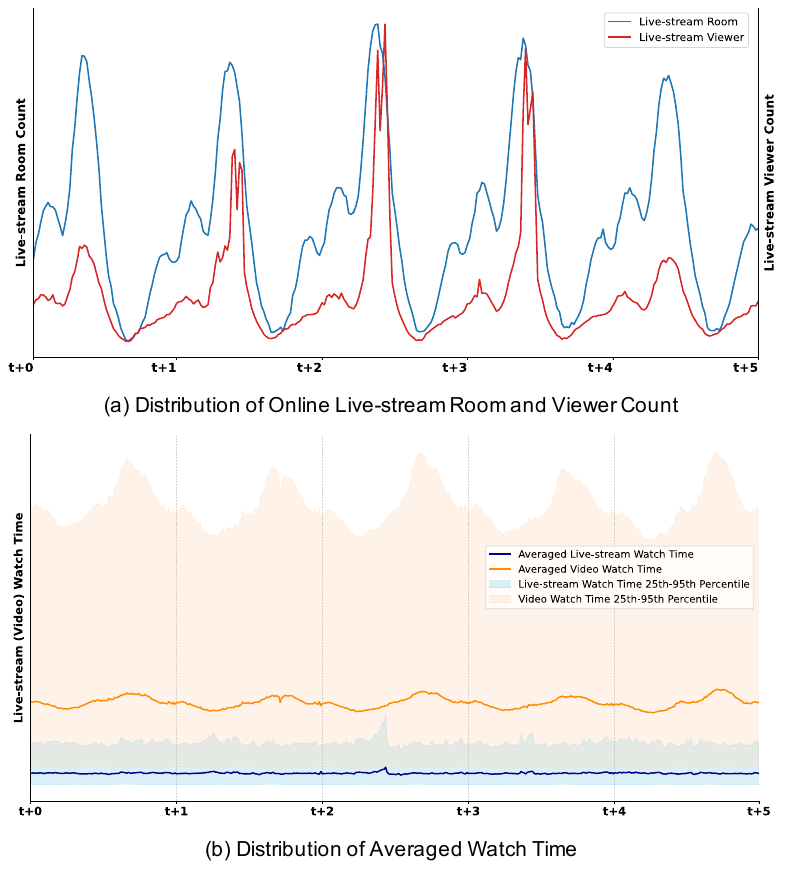}
\caption{Data distributions of online live stream room (or viewer) count, live stream watch time and short video watch time. }
\label{fig:2}
\end{figure}

% \begin{figure}[h!]
% \centering
% \includegraphics[width=0.48\textwidth]{Fig_training_curves.pdf}
% \caption{Training Curves of SL-MGAC and SL-MGAC-sep.}
% \label{fig:7}
% \end{figure}

% \begin{figure}[htbp]
% \centering
% \includegraphics[width=0.48\textwidth]{Fig_ab_live_exposure.pdf}
% \caption{Adaptive adjustments of live stream exposure during peak hours and low-traffic periods.}
% \label{fig:8}
% \end{figure}

\begin{figure}[h!]
\centering
\includegraphics[width=0.8\textwidth]{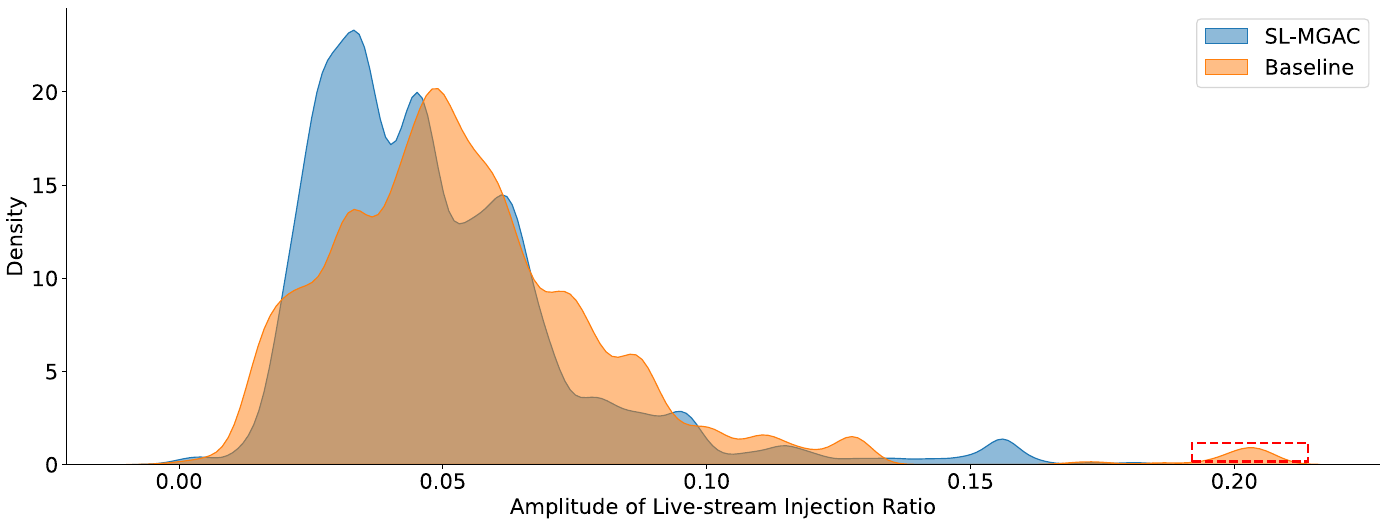}
\caption{The amplitude distribution of live stream injection ratio. }
\label{fig:9}
\end{figure}

\section{More Ablation Studies}
\label{Appdenx: D}
We conduct further online AB experiments to demonstrate the performance of the SL-MGAC policy at different values of $\lambda$ in Eq \ref{eq:8}, as shown in Table \ref{table:5}. It indicates that when $\lambda > 0$, the live stream allocation policy of SL-MAGC behaves more conservatively to balance live stream watch time and video watch time, and finally optimize user's long-term engagement and retention. However, when $\lambda < 0$, the SL-MAGC behaves more greedily to allocate live streams to the video feed. Specifically, we find that the SL-MAGC allocates more live streams to the requests that occur earlier in a session due to the position bias on the reward, which would interrupt the user interest throughout the session and hence affect user retention. 

\begin{table*}[htbp]
\centering
\begin{tabular}{ccccccc}
\hline
$\lambda$ & live stream DAU & live stream Watch Time & Video Watch Time & App Usage Duration & User Retention \\
\hline
0.2 & -1.245\% & -3.567\% & +0.215\% & +0.123\% & +0.106\% \\ 
0.1 & +2.616\% & +7.431\% &+0.197\% & +0.121\% & +0.086\% \\
-0.1 & +4.123\% & +6.564\% & -0.248\% & -0.362\% & -0.112\% \\    
-0.2 & +5.876\% & +7.215\% & -0.296\% & -0.445\% & -0.159\% \\
\hline
\end{tabular}
\caption{Online AB Test Performance of SL-MGAC under different values of $\lambda$.}
\label{table:5}
\end{table*}

\end{document}